\documentclass[11pt, oneside]{article}
\usepackage{authblk}
\usepackage{geometry}
\usepackage{graphicx}
\usepackage{amssymb}
\usepackage{amsmath}
\usepackage{url}
\usepackage{comment}
\usepackage{booktabs}
\usepackage{setspace}
\usepackage[authoryear, round]{natbib}
\usepackage{lineno}
\PassOptionsToPackage{hyphens}{url}
\usepackage{hyperref}
\usepackage{natbib}

\makeatletter

\author[1]{Jackson Gold}
\author[2]{Maria Cuellar}

\affil[1]{University of Pennsylvania School of Engineering and Applied Sciences}
\affil[2]{University of Pennsylvania Department of Criminology and Statistics}

\title{How Often are Fingerprints Repeated in the Population? Expanding on Evidence from AI With the Birthday Paradox}

\begin{document}

\maketitle

\begin{abstract}

The assumption of fingerprint uniqueness is foundational in forensic science and central to criminal identification practices. However, empirical evidence supporting this assumption is limited, and recent findings from artificial intelligence challenge its validity. This paper uses a probabilistic approach to examine whether fingerprint patterns remain unique across large populations. We do this by drawing on Francis Galton’s 1892 argument and applying the birthday paradox to estimate the probability of fingerprint repetition. Our findings indicate that there is a 50\% probability of coincidental fingerprint matches in populations of 14 million, rising to near certainty at 40 million, which contradicts the traditional view of fingerprints as unique identifiers. We introduce the concept of Random Overlap Probability (ROP) to assess the likelihood of fingerprint repetition within specific population sizes. We recommend a shift toward probabilistic models for fingerprint comparisons that account for the likelihood of pattern repetition. This approach could strengthen the reliability and fairness of fingerprint comparisons in the criminal justice system.

\end{abstract}

\section{Introduction}

\subsection{Fingerprint uniqueness is commonly assumed}

The assumption that fingerprints are unique to each individual has long been a foundational assumption of forensic science and criminal investigations \cite{nas2009}, \cite{cole2009forensics}. This fact is required by comparison methods. When examiners conclude that a pair of fingerprints are an ``identification'' without further qualification, this is only useful if fingerprints are unique. Otherwise, an identification only reduces the pool of potential contributors who all have matching fingerprints to an unknown size, a fact that does not have a high probative value on its own.

However, the scientific basis for uniqueness is more tenuous than often presumed by lay individuals and forensic examiners. Indeed, uniqueness is just a common assumption, and it has not been established (\cite{nas2009}, \cite{pcast}, \cite{aaasreport}). In fact, recent research \citep{guo2024unveiling} using artificial intelligence provides empirical evidence that fingerprints coming from two different fingers are indistinguishable from one another. Specifically, they find that fingerprints from different fingers of the same person share very strong similarities, with high confidence.

This article expands on the research from artificial intelligence. It asks, how large does a population have to be before a fingerprint repeats across different individuals? We combine the birthday paradox with Francis Galton’s 1892 experiment on fingerprints to answer this question. 

Answering the question of whether fingerprints are unique is critical for several reasons. Forensic fingerprint comparisons are widely used in criminal cases to identify suspects in crime scenes. However, if the assumption of uniqueness is flawed, the reliability of fingerprint evidence is compromised \citep{cole2009forensics, cole13long}. This has significant legal implications, as fingerprint evidence has often been presented as irrefutable \citep{pcast, kadane2018certainty}. Moreover, a more nuanced understanding of fingerprint patterns and their variability could encourage forensic science to adopt probabilistic approaches, rather than categorical claims of identity, which may improve the accuracy and scientific validity of identifications.

\section{Background}

\subsection{Uniqueness is not clearly defined and has no empirical proof}

The uniqueness of fingerprints is an ill-defined concept. Does uniqueness mean that every person who has ever lived (and ever will live) has different fingerprints? Is it the finger that is unique, or the print? Fingers are all different from each other at the microscopic level, but for the purposes of forensic science we need the latent prints left on objects to be unique, not the fingers themselves. Furthermore, crime scene fingerprints are often partial, distorted, and potentially degraded representations of the skin. Are distorted fingerprints, or those with limited information about the finger, also unique? Are prints left by the same finger always the same? Can examiners reliably discern between two different-source prints? 

Even with a favorable definition of uniqueness for fingerprint examination---such as the idea that high-quality prints left by different fingers on objects are all distinct, consistent over time, show minimal variability across marks made by each finger, and examiners can compare these reliably---there remains no conclusive evidence that such uniqueness actually exists. 

According to the \citep{nas2009}, none of the variabilities of features across a population of fingers or of repeated impressions left by the same finger has been characterized, quantified, or compared. There is some scientific evidence supporting the presumption that friction ridge patterns are unique to each person and persist unchanged throughout a lifetime (\cite{cummins1961finger}, \cite{hale1952morphogenesis}, \cite{holt1968genetics}, \cite{montagna1974}, and \cite{raser2005noise}). However, this research does not provide empirical evidence for the uniqueness of fingerprints \citep{nas2009}. Thus follows a natural question: How often do fingerprints repeat across individuals? Research from over 100 years ago addresses this question.

\subsection{Recent results from artificial intelligence show fingerprints are shared across fingers}

\cite{guo2024unveiling} proved that fingerprints from different fingers of the same person share strong similarities with a confidence of 99.99\%. They did this by using a deep contrastive network. A contrastive network is a paradigm for unsupervised learning that uses techniques from computer vision to vectorize images and then compare them to other training data to find data points that are most and least similar. Unsupervised learning refers to the machine learning problem where one wants to build a model to predict whether a data point will be similar to a specific group of other data points, without using labelled data. 

\cite{guo2024unveiling} asks, does this print fall into the set of prints that match another person? This was achieved here by embedding approximately 3000 fingerprints into a vector form. This vectorization was done via vision algorithms like binarization, orientations, and others (see the features in Figure 2B of the article). Then, the vision techniques used to match the different attributes sought in the prints was selected. A black-box approach with unique methods to highlight different elements, like the minutiae versus the ridges, was used with computer vision to use for vectorization. They, vectors for each image  were generated and compared against each other's print (the authors often refer to the number of pairs instead of the number of prints) using the Euclidean distance between their respective vectors. The authors used these comparisons to readjust the vectors which allowed them to group them by similarity eventually and then make predictions. The name ``contrastive'' comes from the comparison step, where each vector is contrasted against each others in order to find similarities and groups within the data.

In the results section of the article, \cite{guo2024unveiling} mention that they do their initial analysis of the binary patterns, ridge orientation, ridge density, and minutiae as their criteria. This approach aligns very closely with what Galton had proposed and what is common for a human fingerprint analysis to take into account when performing an analysis. We recall that Galton’s method leaned towards the idea that minutiae were critical for distinguishing fingerprints and that the ridge patterns (loops versus whorls, etc.) were significantly less relevant. However, the researchers found the opposite. They concluded that ``ridge orientation maps perform almost as well as the binarized and original images—this suggests that most of the cross-finger similarity can actually be explained by ridge orientation,'' where the binarized versions were used as a baseline to test the above-mentioned features. 

The researchers found that judging only by the minutiae was practically akin to guessing. The researchers tested the effectiveness of each feature individually, which, while interesting, is simplistic, given the holistic effects of their regularization method (that both orientation and minutiae should be taken into account). They claim that minutiae are practically irrelevant to fingerprint matching. Regardless, it is doubtful that any of the top fingerprint analyses or algorithms could accurately match two fingerprints with just the location of minutiae. 

This research shows empirical evidence that fingerprints from two different fingers can indeed be the same, and thus fingerprints are not unique. But how often will there be repeated fingerprints across different individuals?

\subsection{Francis Galton designed an experiment to find out how often fingerprints are shared across individuals}

As described by \cite{stigler1995galton}, fingerprints as a device for personal identification were not widely used before they were introduced in India in the 1870s by Sir William Herschel. In 1880, Herschel and Henry Faulds brought them to public attention in England as a potential method for identifying criminals. In 1890-95 the use of fingerprints acquired a scientific basis thanks to the work of Francis Galton.

\cite{galton1892fingerprints} was interested in the question of whether one could use fingerprints as a way to identify individuals. He wanted to answer the question of whether ``fingerprints were unique or at least sufficiently distinguishable to be used for evidence.'' \citep{stigler1995galton} To do this, he first studied the heritability of fingerprints (he found there were similarities within families) and racial differences (he found there were no similarities by race). Then, he developed an argument to show that fingerprints are unlikely to be repeated across individuals in the human population. Galton determined the evidential value of fingerprints by designing an experiment described below.

\begin{quote}
In order to break a single fingerprint into components, eh posed the question: if a small square were drpoped onto a fingerprint at random, hiding all the portion of the pattern that lay beneath the square, an an experienced analyst attempted to reconstruct by guesswork the hidden portion based on what was observed outside the mall square, how large should the square be for the probability of a successful guess to be 1/2? From experiment he found that a square with a side about the width of six ridges would do the trick -- actually, from 75 trials Galton estimated that the average chance of a successful guess with a six-ridge square would be about 1/3. He believed that a five ridge square would be nearer to the size sought, but he took the six-ridge square in order to err ``on the safe side.'' A full fingerprint consisted of 24 six-ridge squares, and Galton then claimed, ``These six-ridge-interval squares may thus be regarded as independent units, each of which is equally liable to fall into one or other of two alternative classes, when the surrounding conditions are alone known'' (1892, p. 109). Thus, given that each square was guessed with full knowledge of the surroudning territory, he calculated the change of a successful composite guess at $1/2^{24}$, a value he regarded as an overestimate\dots Galton completed his calculation by assessing the chances that he would guess the correct conditions for reconstructing each square. He took as $1/2^4$ the chance that he would have guessed correctly ``the general course of the ridges adjacent to each square,'' and he estimated the chance that he would have correctly guessed the numbers of ridges entering and leaving each square as $1/2^8$\dots This gave him an overall assessment of the chance that a random fingerprint would match a specified one as $1/2^{24} \times 1/2^{8} \times 1/2^{4}=1/2^{36}$, ``or 1 to about sixty-four thousand millions.''
\end{quote}

Galton's argument, then, is that as the number of people on planet Earth was around 1.6 billion when he was writing his argument, the chance that an individual's fingerprint would be exactly like that of the same finger of any other member of the human race was 1 to 4. To be accepted today, Galton's modeling would require more detail and should rely less on an individual's subjective guesses, but with minor changes, it is correct and conservative \citep{stigler1995galton}.

Galton’s calculation of 1 chance in 64 billion has been mentioned repeatedly following his book's publication \citep{stigler1995galton}, and it has served as the theoretical foundation of fingerprint comparisons, which assume that fingerprints are unique. This argument, in addition to other factors, such as a few dramatically successful cases, led to fingerprint comparisons being widely accepted in the criminal justice system.

Although Galton felt certain that he had proved that it is unlikely that two individuals would have the same fingerprint, this was only given the population of the Earth in 1892. In fact, with his experiment he also proved that it is not impossible that two individuals have the same fingerprint. Indeed, with a large enough population, it is almost guaranteed that different individuals have the same fingerprints. But how large does the population have to be before individuals start sharing fingerprints?

\section{Method} 

\subsection{Expanding on Galton's experiment by using the birthday paradox}

The birthday paradox offers a useful framework for exploring the uniqueness of fingerprints because it addresses a core idea from Francis Galton's original argument. Galton proposed that, even with a large number of people, fingerprints are unlikely to repeat across individuals. The birthday paradox, however, reveals a counterintuitive insight into the probability of repeated patterns in large groups. The paradox shows that in a group of just 23 people, there is already a 50\% chance that two people will share the same birthday. This phenomenon is due to the rapid increase in potential pairings as the group size grows. By analogy, if fingerprint patterns were not as distinctive as assumed, we might similarly expect repeated patterns to emerge sooner than anticipated in large populations. Using the birthday paradox to examine fingerprint patterns is intriguing because it highlights the limitations of relying solely on intuition about uniqueness. It allows us to question whether fingerprints are truly unique across large populations. This approach can help quantify the likelihood of encountering identical fingerprint patterns.

To introduce the birthday paradox, we first define the set $\Omega$ which contains t unique items. We set up the problem as a uniform sampling \textit{with replacement} of $\Omega$. We sample from $\Omega$ a total of p times producing an output set $\Omega'$. We will define B(t,p) as the function that outputs the probability that $\Omega'$ is not comprised of unique items, that is, $\exists \omega_i \in \Omega', \omega_j \in \Omega' \text{ where }  i \neq j \text{ and }\omega_i = \omega_j$. We note that this problem is not very interesting in the case where $p \geq t + 1$ as by the pigeonhole principle \footnote{For natural numbers a and b, if n = ab + 1 objects are distributed among b sets, the pigeonhole principle asserts that at least one of the sets will contain at least a + 1 objects.} it is 1 so we will only consider problems where $p \ll t$.

To approach a closed from solution for B(t,p) we consider the complement which is the probability that $\Omega'$ is unique. We can break this down further by using the multiplication rule and consider that for all of the p trials there exists 1 less possible item from $\Omega$ that could be chosen to keep $\Omega'$ unique. For example if p = 3 then,
\begin{equation}
    B(t, 3) = 1 - \left(\frac{t}{t} * \frac{t - 1}{t} * \frac{t - 2}{t} \right),
\end{equation}
which makes logical sense as during the first trial there is no such element that could break the uniqueness. While during the second there is one such element (the one chosen at trial 1) so the probability that uniqueness is maintained is $\frac{t-1}{t}$ as we select items from $\Omega$ i.i.d. with replacement. We can generalize the B function using this idea to get,
\begin{equation}
    B(t, p) = 1\,-\,\prod_{n\,=\,0}^{p-1}\left(\frac{\,t-n\,}{t}\right).
\end{equation}
Or alternativly if we rearrange the bounds, that is, removing the $\frac{t}{t} = 1 $ term,
\begin{equation}
     B(t, p) = 1\,-\,\prod_{n\,=\,1}^{p}\left(\frac{\,t-n\,+\,1}{t}\right).
\end{equation}

The Birthday Paradox we all know and love is a version of this function where the t parameter is fixed at 365 (assuming no leap years) and the p is varied. The Birthday Paradox gained popularity because of its surprising results. It can be shown that $B(365, 23) \approx 0.5$ which in english means that in a room of 23 people there is a $50\%$ chance that two of them share the same birthday. This surprisingly small result surprises most people until one switches one's perspective to the number of pairs of people in the room. Using the binomial coefficients where we choose 2, we see \citep{h2019discrete},
\begin{equation}
    \binom{n}{k} = \frac{n!}{k!(n - k)!}
\end{equation}
\begin{align}
     \binom{n}{2} & = \frac{n!}{2!(n - 2)!} \nonumber \\
     & =  \frac{n (n - 1)}{2},
\end{align}
which for a $n = 23$ results in 253 pairs of people which makes the $50\%$ figure much more tangible.

\section{Results} 

\subsection{Applying the birthday paradox to fingerprints}

\begin{figure}[ht]
    \centering
    \includegraphics[width=\linewidth]{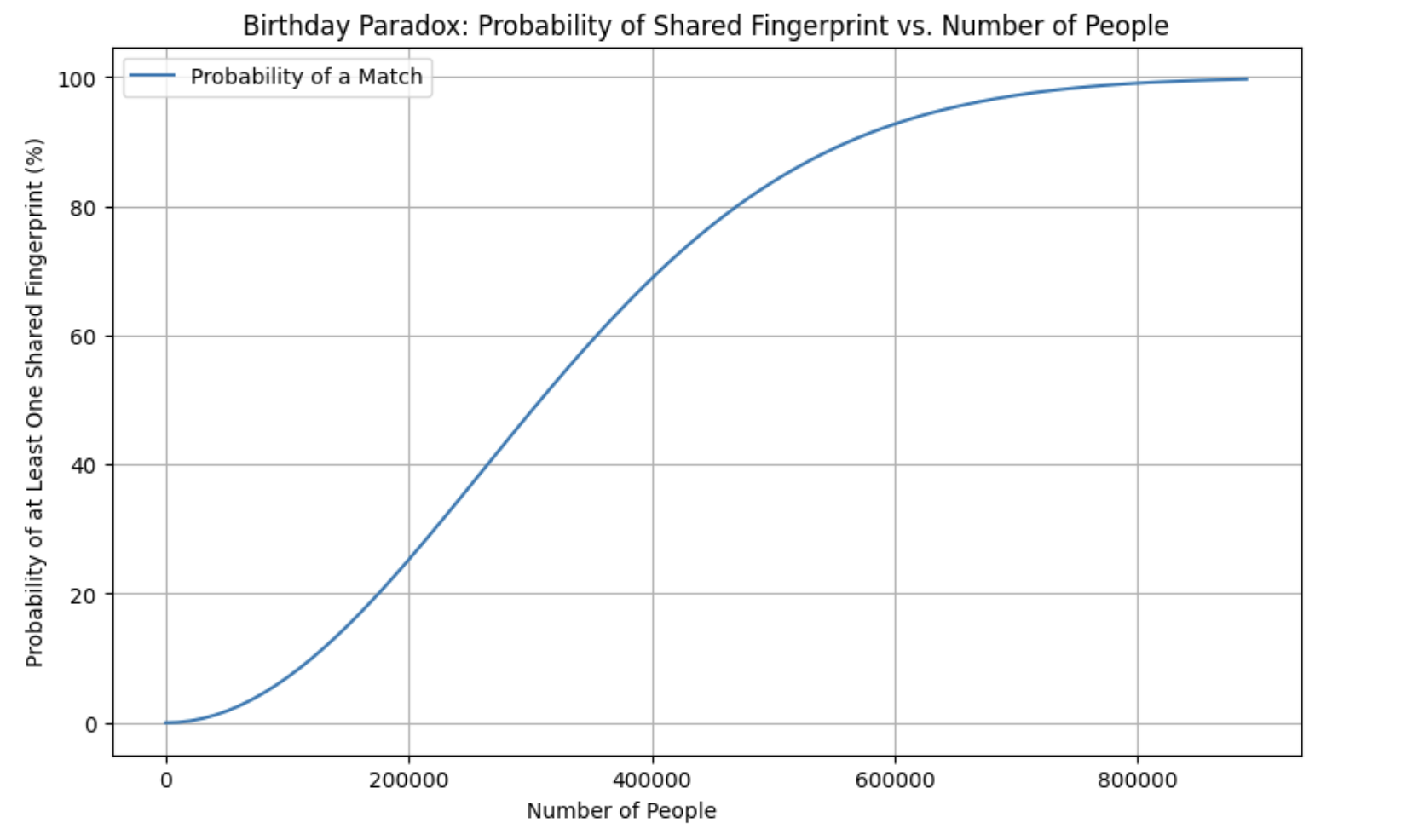}
    \caption{A graph of the result of $B(\gamma, p)$ where $\gamma = 2^{36}$ as Galton's constant and $p$ is varied. Measure in Hundred Thousands}
    \label{fig:gamma-graph}
\end{figure}

Galton came to the conclusion that there are 47 \textit{independent} regions of the print all of which have a binary possibility. From this he determined that there we a total of $2^{47}$ possible fingerprints, assuming that each of the segments of the print are independent. We can now using Galton's total fingerprint approximation as a constant $\gamma$, we can apply the birthday paradox to gain incites on the potential overlap between fingerprints. Using $B(\gamma,p)$ where p is varied we can see the output of B in Figure ~\ref{fig:gamma-graph}.

The results of Figure~\ref{fig:gamma-graph} show that $B(\gamma, 1.4 * 10^7) \approx 0.5$ and $B(\gamma, 4 * 10^7) \approx 0.997$ which correspond to there being a $50\%$ chance of an overlapping fingerprint in a group of 14 million and a nearly $100\%$ chance of an overlap in a group of 40 million. For context of those numbers see Table~\ref{tab:cities}. This table includes a list of cities, their corresponding populations, and the probability that there will be two individuals in this city with indistinguishable fingerprints, according to our application of the birthday paradox.

\begin{table}[ht]
\centering
\begin{tabular}{lcc}
\toprule
City & Population & Random Overlap Probability (\%) \\
\midrule
New York City   & 8,419,600 & $\approx$ 100\% \\
Los Angeles     & 3,980,400 & $\approx$ 100\%\\
Chicago         & 2,746,388 & $\approx$ 100\%\\
Nashville       & 687,788 & 96.80 \% \\
Las Vegas       & 660,929 & 95.83 \% \\
Detroit         & 633,218 & 94.59 \%\\
Baltimore       & 565, 239 & 90.22\% \\
Atlanta         & 510,823 & 85.02 \% \\
Raleigh         & 482,295 & 81.59 \% \\
Miami           & 467,963  & 79.68\% \\
Minneapolis     & 429,606  & 73.89\% \\
Tulsa           & 413,066  & 71.10\% \\
Arlington       & 398,854  & 68.57\% \\
New Orleans     & 376,971  & 64.44\% \\
Wichita         & 397,532  & 68.33\% \\
Cleveland       & 367,991  & 62.67\% \\
Tampa           & 384,959  & 65.98\% \\
Aurora          & 386,261  & 66.23\% \\
Anaheim         & 345,940  & 58.14\% \\
Honolulu        & 345,510  & 58.05\% \\
Lexington       & 322,570  & 53.10\% \\
Anchorage       & 291,247  & 46.05\% \\

\bottomrule
\end{tabular}
\caption{Random Overlap Probabilities for US cities. That is, the result of $B(\gamma, p)$ where p is replaced with the population of the city. Cities with populations larger than those showed here also have a ROP $\approx$ 100\%.}
\label{tab:cities}
\end{table}

\subsection{Defining a Random Overlap Probability}

We define the Random Overlap Probability (R.O.P.) as the probability that there are equal fingerprints among at least two different individuals in a group. That is, $B(\gamma, p)$ where with $p$ being the overlap being the size of the group you can narrow down the print to. Much like the concept of Random Match Probability in DNA analysis, R.O.P. is  a measure to determine the strength of a print's individual uniqueness as a function of the size of the pool of suspects. Table \ref{tab:cities} shows the R.O.P. of cities in the United States. It should be noted that many of the United States' major cities included in the table have their R.O.P. approximated to $100\%$. These include but are not limited to, New York City, Los Angeles, Chicago, and Philadelphia. With this metric, almost no major city police department would be able to, with confidence, conclude the guilt of an individual based exclusively on a latent print.

Consider the following example to illustrate the usage of the R.O.P. Suppose that the Miami-Dade Police Department finds a print and the police is certain that the perpetrator lives in Miami-Dade country. Then, the R.O.P. would be 79.68\%. Another way of understanding the R.O.P. is that it indicates the strength of which a latent print can be used as the sole form of evidence for making a conviction. 

Applying this concept to a real case demonstrats its use. The implementation of the R.O.P. could have been beneficial when retroactively looking at infamous 2004 Brandon Mayfield case\citep{oig2006review}. Had the FBI used the R.O.P. metric where $p$ is the population value of the United States and Spain, the R.O.P. would be 100\%, which would not have been sufficient to arrest Brandon Mayfield, solely based on the evidence of the latent print.

\subsection{Limitations of this analysis}

There are a few factors that could be a source of error for the aforementioned analysis. The most pressing is the value of $\gamma$. While it is conceptually very difficult to arrive at the total number of fingerprints that are theoretically possible Galton's appears to be the best estimate within the literature. Another source of error comes from the fact that we assumed that each section of ones fingerprint is unique. This assumption was commonly though to be true but a recent study \citep{guo2024unveiling} that showed with contrastive twin neural networks that finger prints on the same hand of a person statistically significantly similar to each other. This may imply that not all segment of a fingerprint which could dramatically decrease the true value of $\gamma$ as it applies to fingerprints. Finally, we only considered the question of whether fingerprints repeat across individuals, and not whether they repeat within individuals. We did not include the fact that individuals have ten fingerprints as part of our calculations.

\subsection{Theoretical upper bound for global uniqueness}

We will now introduce the notion of a $\Gamma_x$ constant. The $\Gamma_x$ constant is the of t in B(t,p) for which when we set p to the global population $\phi = 8.2 * 10^9$, the value of $B(\Gamma_x, \phi) = x * 100$. That is $\Gamma_x$ is the upper bound on the number of unique item such that that is a $x\%$ chance of an overlap among the world. While this could be done by hand, the close form solution it a bit tedious. However, it can be done by gradient descent on $B$ which is effectively convex. For sake of optimization we can make use of Stirling's approximation for factorials or cumulative summation formulas to handle the large ranges efficiently. It would be much to computationally expensive to calculate all of the $\Gamma_x$ values but $\Gamma_{50}$ is a good benchmark and was found to be:
\begin{equation*}
    B(\Gamma_{25}, \phi) \approx 1.1686 \times 10^{20}\\
\end{equation*}
\begin{equation*}
    B(\Gamma_{50}, \phi) \approx 4.8502 \times 10^{19}
\end{equation*}
\begin{equation*}
    B(\Gamma_{75}, \phi) \approx 2.4252 \times 10^{19}
\end{equation*}

In simple terms $\Gamma_x$ is the theoretically number of fingerprints such that uniqueness is guaranteed over the entirely world population with a probability of $x\%$. For instance, if there were $2.4252 \times 10^{19}$ fingerprints in the world, then there is a 75\% change that they would be unique. This can be extrapolated to other forms of analysis with known upper bounds to show a relative semblance of uniqueness. 

\section{Recommendation for improving fingerprint comparisons}

Triers of fact should know about the nature of fingerprints and the information contained in them. It is important to inform them not just about the expert's conclusion, but also give them some background about fingerprints themselves. They should know that fingerprints are not unique, and in fact are very likely to repeat across different individuals, especially in large populations. 

One way to move towards a different method that does not assume uniqueness is by using a probabilistic framework (a suggestion that has been made by many, including \cite{champod2001probabilistic}, \cite{kaye2013beyond}, \cite{cole13long}, and \cite{neumann2012quantifying}). A likelihood ratio approach has been recommended as a way to give  probabilistic conclusions. While adopting a probabilistic model could be beneficial, even a more moderate re-evaluation would improve forensic practices. Kaye found that, although the probability of finding identical fingerprints among two random individuals may not be negligible, the probability that a specific individual shares a fingerprint with another specific person remains exceedingly low. It is sufficient to move away from the categorical mindset that fingerprints are an absolute identifier, and instead recognize the nuanced probabilities involved. 

Additionally, it is important to present fact finders with the results from methods that have been shown to be foundationally valid and admissible in court. This should be done by following the admissibility standards from the law, such as the Federal Rules of Evidence 702, the Frye standard, and the Daubert standard. For decades, fingerprint comparisons have been presented as infallible to the public \citep{kadane2018certainty}. We now know that fingerprint comparisons are not infallible and in fact have higher error rates that lay individuals expect \citep{pcast}. Triers of fact should hear about the accuracy and consistency of fingerprint comparisons (e.g., \citep{ulery2011accuracy}). This might be difficult for experts to communicate to lay people, since \cite{kadane2018certainty} showed that jurors do not value the evidence any more or less when the examiner uses very strong language to indicate that the defendant is the source of the print versus weaker source identification language. In other words, regardless of whether the pseudo-jurors were told that ``Mr.~Johnson is the source'' or I ``Cannot exclude Mr.~Johnson,'' they heard the same thing: that the fingerprints matched. Nevertheless, it is crucial for the triers of fact to understand the information about the accuracy and consistency of the method. More research needs to be done to learn about jurors' understanding of probabilistic assertions about forensic comparison disciplines.

\section{Conclusion}

A recent result from artificial intelligence \citep{guo2024unveiling} showed that fingerprints from different fingers can be the same, ``with 99.99\% confidence.'' We take this line of research further and find the answer to the question: In how large a population will there be repeated fingerprints across different individuals, with high probability? We use a combination of an experimental, historical approach from \cite{galton1892fingerprints} and combine it with a theoretical approach, the birthday paradox. 

We find that there is a 50\% probability of coincidental fingerprint matches in a population of 14 million, with near certainty at 40 million. This challenges the traditional assumption that fingerprints are unique, and it provides a quantification of how often are fingerprints repeated, via the Random Overlap Probability. The Random Overlap Probability is the probability that there are equal fingerprints among at least two different individuals in a group. Although fingerprints contain complex information, this is not sufficient to guarantee uniqueness across even moderately sized populations, such as the population of certain cities (e.g., Chicago, Los Angeles, and New York City). As population sizes increase, so does the likelihood of coincidental similarities between fingerprint patterns.

This finding has implications for forensic practice and legal proceedings that depend on fingerprint evidence. Since fingerprint patterns are not unique, the reliability of categorical conclusions, particularly in cases where such evidence is the primary or sole link to a suspect, needs to provide a measure of uncertainty (e.g., ``the probability of another unrelated individual having the same fingerprint in a city of this population is $x$''). An even better approach would be to switch from categorical to probabilistic reporting.

Without this measure of uncertainty, the prosecutor risks suffering from the prosecutor's fallacy. The prosecutor's fallacy in this case is stating that, since the crime scene print and a suspect's print are an identification, it is likely that the suspect must have committed the crime. But, in fact, there might be several individuals, even within the city, with the same prints, and thus the probability that the suspect is innocent is actually high.

Forensic science would benefit from empirically validated standards that address the probability of pattern repetition. Probabilistic ways of comparing fingerprints should be used, and a measure of uncertainty should be reported along with a comparison result. By questioning the longstanding assumption of uniqueness, this article supports a reevaluation of fingerprint identification methods, and it contributes to a more reliable criminal justice system.

\bibliographystyle{plainnat}
\bibliography{references.bib}

\end{document}